\documentclass{jpp}

\usepackage{graphicx,bm}
\usepackage{natbib}
\usepackage{color}

\newcommand{\bec}[1]{\mbox{\boldmath $ #1$}}
\newcommand{\EQ}{\begin{equation}}
\newcommand{\EN}{\end{equation}}
\newcommand{\EQA}{\begin{eqnarray}}
\newcommand{\ENA}{\end{eqnarray}}

\newcommand{\meanrho}{\overline{\rho}}

{}
{}
{}

\newcommand{\meanAAA}{\overline{\mathsf{A}}}

{}
{}
{}
{}
{}
{}
{}
{}
\newcommand{\meanBB}{\overline{\mbox{\boldmath $B$}}{}}{}
{}
{}
{}
{}
{}
{}
{}
{}

{}
{}
{}

\newcommand{\meanB}{\overline{B}}

{}

{}
{}

\def\half{{\textstyle{1\over2}}}

\title[Nonlinear mean-field dynamo and predictions of solar activity]{Nonlinear mean-field dynamo and prediction of solar activity}

\author[N. Safiullin, N. Kleeorin, S. Porshnev, I. Rogachevskii and A. Ruzmaikin]%
{N.\ns S\ls A\ls F\ls I\ls U\ls L\ls L\ls I\ls N$^{1}$,
N.\ns K\ls L\ls E\ls E\ls O\ls R\ls I\ls N$^{2,3}$,
S.\ns P\ls O\ls R\ls S\ls H\ls N\ls E\ls V$^{1}$,
I.\ns R\ls O\ls G\ls A\ls C\ls H\ls E\ls V\ls S\ls K\ls I\ls I$^{2,3}$,
  \thanks{Email address for correspondence: gary@bgu.ac.il}
  \and
A.\ns R\ls U\ls Z\ls M\ls A\ls I\ls K\ls I\ls N$^{4}$
}
\affiliation{
$^{1}$Department of Information Technology and Automation,
Ural Federal University, 19 Mira str., 620002 Ekaterinburg, Russia
\\
$^{2}$Department of Mechanical Engineering, Ben-Gurion University of
the Negev, P. O. Box 653, 84105 Beer-Sheva, Israel
\\
$^{3}$Nordita, KTH Royal Institute of Technology
and Stockholm University, Roslagstullsbacken 23,
10691 Stockholm, Sweden
\\
$^{4}$Jet Propulsion Laboratory, California Institute of Technology, 4800 Oak Grove Drive, Pasadena, CA 91109, USA
}

\date{\today; revised ; accepted  }
\begin{document}

\maketitle

%\preprint{NORDITA-2017-134}

\begin{abstract}
We apply a nonlinear mean-field dynamo model which includes a budget equation for the dynamics of Wolf numbers to predict solar activity.
This dynamo model takes into account the algebraic and dynamic nonlinearities of the alpha effect, where the equation for the dynamic nonlinearity is derived from the conservation law for the magnetic helicity.
The budget equation for the evolution of the Wolf number is based on a formation mechanism   of sunspots related to the negative effective magnetic pressure instability.
This instability redistributes the magnetic flux produced by the mean-field dynamo.
To predict solar activity on the time scale of one month we use a method based on a combination of the numerical solution of the nonlinear mean-field dynamo equations and the artificial neural network.
A comparison of the results of the prediction of the solar activity with the observed Wolf numbers demonstrates a good agreement between the forecast and observations.
\end{abstract}

\section{Introduction}

Since formulation of the mean-field dynamo approach in a seminal paper by Steenbeck, Krause
and R\"{a}dler in 1966 \citep{Steenbeck(1966),Krause(1980),Roberts(1971)}, the theories of solar magnetic fields have been actively developing during last 50 years \citep{Moffatt(1978),Parker(1979),Krause(1980),Zeldovich(1983),Ossendrijver(2003),Brandenburg(2005),Ruediger(2013)}.
Well-known point of view on origin of the large-scale solar magnetic field is that the field is generated in the solar convective zone by a combine action of helical convective turbulent motions and large-scale non-uniform rotation (so called $\alpha\Omega$ or $\alpha^2\Omega$ mean-field dynamo). This large-scale magnetic field causes the 11-year solar cyclic activity.

However, the observed solar activity is associated with strongly concentrated magnetic fields as sunspots with the characteristic spatial scale of the order of solar super-granulation (about $10^4$ km). On the other hand, the mean-field dynamo generates smooth large-scale magnetic fields with the characteristic scale of the order of the solar radius (about $10^6$ km). So it was not clear many years, how to relate the mean-field dynamo with the sunspots.
One of the suggested mechanism of magnetic spots formation is the magnetic buoyancy instability suggested by Parker in 1966 \citep{Parker(1966),Parker(1979),Priest(1982)}. This instability is excited when the characteristic scale of original magnetic field variation is smaller than the density stratification height. Therefore, in strongly density stratified convective zone where the density varies in radial direction by 7 orders of magnitude, this instability is only excited when the initial magnetic field is already strongly non-uniform.

Another mechanism of magnetic spot formation is related to the negative effective magnetic pressure instability (NEMPI), which can be excited even for uniform initial large-scale magnetic field \citep{Kleeorin(1989),Kleeorin(1990)}.
The mechanism of this instability is based on the suppression
of total (hydrodynamic and magnetic) turbulent pressure
by large-scale magnetic field.
NEMPI can be understood as a negative contribution
of turbulence to the effective mean magnetic pressure (the sum
of non-turbulent and turbulent contributions).
At large fluid and magnetic Reynolds
numbers this turbulent contribution becomes large and a large-scale
instability  can be excited, redistributing the magnetic flux produced
by the mean-field dynamo.
NEMPI has been studied analytically
using the mean-field approach \citep{Kleeorin(1994),Kleeorin(1993),Kleeorin(1996),Rogachevskii(2007)}
and numerically using mean-field simulations \citep{Brandenburg(2010),Brandenburg(2014),Brandenburg(2016),Kemel(2012),Kemel(2013)}, large-eddy simulations \citep{Kapyla(2012),Kapyla(2016)} and direct numerical simulations \citep{Brandenburg(2011),Brandenburg(2012),Brandenburg(2013),Mitra(2014),Warnecke(2013),Warnecke(2016),Jabbari(2016)}.

Predictions of solar activity is a subject of active investigations
where various methods including the mean-field dynamo models have been used \citep{Dikpati(2006),Choudhuri(2007),Kane(2007),Bushby(2007),Obridko(2008),DeJager(2009),Kitiashvili(2011),Tlatov(2009),Tlatov(2015),Pesnell(2012),Usoskin(2017)}.
However, the improving of the solar activity forecast
is still a subject of numerous discussions.

In the present study we apply a nonlinear mean-field dynamo model and a budget equation for the dynamics of Wolf numbers \citep{Kleeorin(2016)} to predict solar activity.
This budget equation is related to a mechanism of formation of sunspots based on NEMPI.
The dynamo model includes the algebraic and dynamic quenching of the alpha effect.
To predict solar activity on the time scale of one month we use a method based on a combination of the numerical solution of the nonlinear mean-field dynamo equations \citep{Kleeorin(2016)} and the artificial neural network approach \citep{Hagan(2016),Conway(1998),Fessant(1996)}.

\section{Nonlinear dynamo model}
\label{model}

To study nonlinear evolution of the large-scale magnetic field, we use
the induction equation in spherical coordinates $(r, \theta, \phi)$ for an
axisymmetric mean magnetic field, $ \meanBB = \meanB_\phi
\bec{e}_{\phi} + \bec{\nabla} {\bf \times} (\meanAAA \bec{e}_{\phi})$.
We investigate the dynamo action in a thin convective shell.
To take into account strong variation of the plasma density in the radial direction, we
average the equations for the mean toroidal field $\meanB_\phi$ and the magnetic potential $\meanAAA$ of the mean poloidal field over the depth of the
convective shell, so that all quantities are functions of
colatitude $\theta$.
We neglect the curvature of the convective shell
and replace it by a flat slab.
These simplifications yield the
non-dimensional mean-field dynamo equations \citep{Kleeorin(2003)}:
\begin{eqnarray}
{\partial \meanB_\phi \over \partial t} &=& G D \sin \theta {\partial
\meanAAA \over \partial \theta} + {\partial ^2 \meanB_\phi \over \partial
\theta^2} - \mu ^2 \meanB_\phi ,
\label{M1}\\
{\partial \meanAAA \over \partial t} &=& \alpha \meanB_\phi + {\partial^2 \meanAAA
\over \partial \theta^2} - \mu^2 \meanAAA ,
\label{M2}
\end{eqnarray}
where the terms, $-\mu^2 \meanB_\phi$ and $-\mu^2
\meanAAA$, in Eqs.~(\ref{M1}) and~(\ref{M2}) describe turbulent diffusion
of the mean magnetic field in the radial direction,
the parameter $G =\partial \Omega / \partial r$ determines the differential rotation,
and $D$ is the dynamo number defined below.
The parameter $\mu$ is determined by the equation:
\begin{eqnarray}
\int_{2/3}^{1} {\partial^2 \meanB_\phi \over \partial r^2} \,dr
= - {\mu^2 \meanB_\phi \over 3} .
\label{M3}
\end{eqnarray}
In Eqs.~(\ref{M1})--(\ref{M3}) the length is measured in
units of radius $R_\odot$, time is measured in units of the turbulent
magnetic diffusion time $R_\odot^2 / \eta_{_{T}}$, the differential
rotation $\delta\Omega$ is measured in units of the maximal value
of the angular velocity $\Omega$,
and $\alpha $ is measured in units of the maximum value of the
kinetic part of the $ \alpha $-effect.
The toroidal mean magnetic field, $\meanB_\phi$ is measured in the units of the equipartition field $\meanB_{\rm eq} = u_0 \sqrt{4 \pi \meanrho_\ast}$, and the vector potential of the mean poloidal field $\meanAAA$ is measured in units of $R_{\alpha} R_\odot \meanB_{\rm eq}$.
The density $\meanrho$ is normalized by its value $\meanrho_\ast$ at the bottom of the convective zone, and the integral scale of the turbulent motions $\ell_0$ and turbulent
velocity $u_0$ at the scale $\ell_0$ are measured in units of their
maximum values through the convective region. The
magnetic Reynolds number ${\rm Rm}=\ell_0 u_0/\eta$
is defined using these maximal values, and the turbulent magnetic diffusivity
is $\eta_{_{T}}=\ell_0 u_0 / 3$. Here $\eta$ is the magnetic diffusion coefficient due to electrical conductivity of plasma.
The dynamo number is defined as $D = R_\alpha
R_\omega$, where $R_{\alpha} = \alpha_0 R_\odot / \eta_{_{T}}$ and
$R_\omega = (\delta \Omega) \, R_\odot^2 / \eta_{_{T}}$.
Equations~(\ref{M1}) and~(\ref{M2})
describe the dynamo waves propagating from the central
latitudes towards the equator when the dynamo number is negative.
The radius $r$ varies from $2/3$ to $1$
inside the convective shell, so that
the value $\mu=3$ corresponds to a convective
zone with a thickness of about 1/3 of the radius.

The total $\alpha$ effect is defined as the sum of the kinetic,
$\alpha^v= \chi^v \phi_{v}(\meanB)$, and magnetic, $\alpha^m= \chi^c \phi_{m}(\meanB)$, parts:
\begin{eqnarray}
\alpha (r, \theta) = \chi^v \phi_{v}(\meanB) + \chi^c \phi_{m}(\meanB) ,
\label{M4}
\end{eqnarray}
where $\chi^v = - (\tau_0 /3) \, \overline{\bec{u}\cdot(\bec{\nabla} {\bf \times} \bec{u})}$, $\, \chi^{c} = (\tau_0 / 12 \pi \meanrho)\, \overline{{\bm b} \cdot (\bec{\nabla} {\bf \times} {\bm b})}$ and $\tau_0$ is the correlation time of the turbulent velocity field.
We adopt the standard profile of the kinetic part of the $\alpha$ effect:
$\alpha(\theta)=\alpha_0 \sin^3 \theta \cos \theta$.
The magnetic part of the $\alpha$ effect \citep{Pouquet(1976)} and density of the magnetic helicity are related to the density of the current helicity $\overline{{\bm b} \cdot (\bec{\nabla} {\bf \times} {\bm b})}$  in the approximation of weakly inhomogeneous turbulent convection \citep{Kleeorin(1999)}.
The quenching functions $\phi_{v}(\meanB)$ and $\phi_{m}(\meanB)$ in
Eq.~(\ref{M4}) are given by:
\begin{eqnarray}
\phi_{v}(\meanB) &=& (1/7) [4 \phi_{m}(\meanB) + 3 L(\meanB)] ,
\label{M5} \\
\phi_{m}(\meanB) &=& {3 \over {8\meanB^2}} [1 - \arctan (\sqrt{8} \meanB) /
\sqrt{8} \meanB],
\label{M6}
\end{eqnarray}
\citep{Rogachevskii(2000),Rogachevskii(2001)}, where $ L(\meanB) = 1 - 16
\meanB^{2} + 128 \meanB^{4} \ln (1 + 1/(8\meanB^2))$.
The quenching functions have the following asymptotics:
$\phi_{v}(\meanB) = 1-(48/5)\meanB^2$ and $\phi_{m}(\meanB) = 1-(24/5)\meanB^2$
for weak magnetic field, $\meanB \ll 1$,
while $\phi_{v}(\meanB) = 1/(4\meanB^2)$ and $\phi_{m}(\meanB) = 3/(8\meanB^2)$ for
strong magnetic field, $\meanB \gg 1$, where
$ \chi^v $ and $ \chi^c$ are measured in units of maximal
value of the $\alpha$-effect.
The function $\phi_{v}$ describes the algebraic quenching of the
kinetic part of the $\alpha $ effect that
is caused by the effects of the mean magnetic field
on the electromotive force \citep{Rogachevskii(2000),Rogachevskii(2001),Rogachevskii(2004)}.

We average Eq.~(\ref{M4}) over the depth of the
convective zone, so that the first term in the averaged equation is
determined by the values taken at the middle part of
the convective zone, while in the second term there is a
phenomenological parameter $\sigma$:
\begin{eqnarray}
\alpha (\theta) = \chi^v \phi_{v}(\meanB) + \sigma \chi^c \phi_{m}(\meanB) ,
\quad \quad
\sigma = \int \left({\meanrho(r)\over \meanrho_\ast} \right)^{-1} \, dr,
\label{M7}
\end{eqnarray}
\citep{Zhang(2012),Kleeorin(2016)},
where the densities of the helicities and quenching functions are 
associated with a middle part of the convective zone. The parameter $\sigma > 1$ is a free parameter.

The magnetic part $\alpha^m$ of the $\alpha $ effect is based on two
nonlinearities: the algebraic quenching, given by the function
$\phi_{m}(\meanB)$
\citep{Field(1999),Rogachevskii(2000),Rogachevskii(2001)} and the dynamic nonlinearity.
The function $\chi^c(\meanBB)$ is determined by a dynamical
equation that is derived using the conservation law for magnetic
helicity \citep{Kleeorin(1982),Gruzinov(1994),Kleeorin(1995)}:
\begin{eqnarray}
{\partial \chi^{c} \over \partial t} + \bec{\nabla} \cdot \bec{\Phi} + {\chi^c \over T} = -{1
\over 9 \pi \, \eta_{_{T}} \, \meanrho_\ast} \, (\bec{\cal E} {\bf \cdot}
\meanBB) ,
\label{M8}
\end{eqnarray}
where $ \bec{\Phi} = -\kappa_{_{T}} \bec{\nabla} \chi^c$
is the turbulent diffusion flux of the density of the magnetic helicity
\citep{Kleeorin(1999),Kleeorin(2000),Kleeorin(2003),Blackman(2000),Brandenburg(2005)},
$\kappa_{_{T}}$ is the coefficient of the turbulent diffusion,
$T = \ell_0^2 / \eta$ is the relaxation time of magnetic helicity and $\bec{\cal E}$ is the mean electromotive force.
Since the total magnetic helicity is conserved,
the increase of the density of the the large-scale magnetic helicity due to the dynamo action,
should be compensated by the decrease of the density of the small-scale magnetic helicity.
The compensation mechanisms include the dissipation and transport of the density of the magnetic helicity.
The dynamical equation~(\ref{M8}) for the function
$\chi^c(\meanBB)$ in non-dimensional form reads
\begin{eqnarray}
&& {\partial \chi^c \over \partial t} + \left(T^{-1} + \kappa_{_{T}}
\mu^2\right)\chi^c = 2\left({\partial
\meanAAA \over \partial \theta} {\partial \meanB_\phi \over \partial \theta} + \mu^2 \meanAAA \, \meanB_\phi\right)
- {\alpha \over\xi} \meanB^2 - {\partial \over \partial \theta} \left(\meanB_\phi {\partial \meanAAA \over \partial \theta} - \kappa_{_{T}} {\partial \chi^c
\over \partial \theta} \right) ,
\nonumber\\
\label{M9}
\end{eqnarray}
\citep{Zhang(2012),Kleeorin(2016)}, where
\begin{eqnarray}
\meanB^2 = \xi \left\{\meanB_\phi^2 + R_\alpha^2 \left[\mu^2 \meanAAA^2 + \left( {\partial \meanAAA \over \partial \theta}\right)^2\right] \right\},
\label{M10}
\end{eqnarray}
and $\xi=2 (\ell_0/R_\odot)^2$.
In derivation of Eqs.~(\ref{M9})--(\ref{M10}), we average Eq.~(\ref{M8}) over the depth
of the convective zone, so that the average value of $T^{-1}$ is
\begin{eqnarray}
T^{-1} = H^{-1} \int T^{-1}(r) \,d r \sim {\Lambda_\ell \, R_\odot^2 \,
\eta \over H \, \ell_0^2 \, \eta_{_{T}}} ,
\label{M11}
\end{eqnarray}
where $H$ is the depth of the convective zone,
$\Lambda_\ell$ is the characteristic scale of variations $\ell_0$, and
$T(r) = (\eta_{_{T}} / R_\odot^{2}) (\ell_0^2 / \eta)$ is the non-dimensional
relaxation time of the density of the magnetic helicity. The values
$\Lambda_\ell , \, \eta, \, \ell_0$ in Eq.~(\ref{M11}) are associated
with the upper part of the convective zone.

In view of observations, an important parameter of the solar activity
is the Wolf number \citep{Gibson(1973),Stix(2012)}, $W= 10 g +f$, where $g$ is the number of sunspot groups and $f$ is the total number of sunspots in the visible part of the sun. This parameter has been measured during three centuries.
Based on the idea of NEMPI, we derive a budget equation
for the surface density of the Wolf number \citep{Kleeorin(2016)}:
\begin{eqnarray}
{\partial \tilde W \over \partial t} = I(t,\theta) - {\tilde W \over \tau_s(\meanB)} .
\label{M12}
\end{eqnarray}
Equation~(\ref{M12}) includes the rate of production of the surface density of the Wolf number, $\tilde W(t,\theta)$, caused by the formation of sunspots:
\begin{eqnarray}
I(t,\theta) = {|\gamma_{\rm inst}| |\meanB-\meanB_{\rm cr}| \over \Phi_s} \Theta(\meanB-\meanB_{\rm cr}) ,
\label{M14}
\end{eqnarray}
and the rate of decay of sunspots, $\tilde W / \tau_s(\meanB)$,
where the decay time, $\tau_s(\meanB)$, of sunspots is discussed below
and $\Theta(x)$ is the $\Theta$ function, defined as $\Theta(x) = 1$ for $x>0$, and $\Theta(x) = 0$ for $x\leq 0$.

The growth rate of NEMPI, $\gamma_{\rm inst}$, is given by \citep{Rogachevskii(2007),losada2012,Brandenburg(2016)}:
\begin{eqnarray}
\gamma_{\rm inst} \approx \left( {2v_{\rm A}^2 k_x^2\over H_\rho^2 k^2}
\left|{d P_{\rm eff} \over d \beta^2}\right| - {4 ({\bm \Omega} \cdot {\bm k})^2
\over {\bm k}^2}\right)^{1/2} - \eta_{_{T}} \left(k^2 + {1 \over (2 H_\rho)^{2}}\right),
\label{M15}
\end{eqnarray}
where $v_{\rm A}=\meanB/\sqrt{4 \pi \meanrho}$ is the mean Alfv\'en speed, ${\bm k}$ is the wave number, ${\bm \Omega}$ is the angular velocity, $P_{\rm eff}=\half\left[1-q_{\rm p}(\beta)\right]\beta^2$ is the effective magnetic pressure, the nonlinear function $q_{\rm p}(\beta)$ is the turbulence contribution to the mean magnetic pressure and $\beta=\meanB/\meanB_{\rm eq}$.
As follows from Eq.~(\ref{M15}), NEMPI is excited in the upper part of the convective zone,
where the Coriolis number ${\rm Co}= 2 \Omega \, \tau_0$ is small.
To determine the source function $I(t,\theta)$ given by Eq.~(\ref{M14}),
we take into account that NEMPI has a threshold, i.e., the instability is excited $(\gamma_{\rm inst}>0$), only when the mean magnetic field is larger than a critical value,
$\meanB>\meanB_{\rm cr}$. This implies that the source $I(t,\theta)$ is proportional to a $\Theta$ function.
The critical value $\meanB_{\rm cr}$ of the mean magnetic field
is given by
\begin{eqnarray}
{\meanB_{\rm cr} \over \meanB_{\rm eq}} = {\ell_0 \over 50 H_\rho} \left[1 + \left({10 \, {\rm Co} \, H_\rho^2 \over \ell_0^2} \right)^2 \right]^{1/2} ,
\label{T20}
\end{eqnarray}
where we use Eq.~(\ref{M15}).
For upper part of the convective zone, this field
$\meanB_{\rm cr} \geq \meanB_{\rm eq} / 50$ is small enough.

The function $I(t,\theta)$ determines the Wolf number variation rate.
The characteristic time of the Wolf number variations is assumed to be identified with the characteristic time for excitation of the instability, $\gamma_{\rm inst}^{-1}$.
When $\gamma_{\rm inst}<0$, the rate of production, $I(t,\theta)$, vanishes.
This implies that the function $I(t,\theta) \propto |\gamma_{\rm inst}| \, \Theta(\meanB-\meanB_{\rm cr})$.
The production term, $I(t,\theta)$, is also proportional to the maximum number of sunspots per unit area, that can be estimated as $\sim |\meanB-\meanB_{\rm cr}| /\Phi_s$, where $|\meanB-\meanB_{\rm cr}|$ is the magnetic flux per unit area that contributes to the sunspot formation and $\Phi_s$ is the magnetic flux inside a magnetic spot.

The decay of sunspots during the nonlinear stage of NEMPI, is described by the relaxation term, $-\tilde W /\tau_s(\meanB)$.
The decay time $\tau_s(\meanB)$ varies from several weeks to a couple of month, while the solar cycle period is about 11 years. Therefore, to determine the surface density of the Wolf number, we can use the steady-state solution of Eq.~(\ref{M12}): $\tilde W = \tau_s(\meanB) \,I(t,\theta)$. The Wolf number is defined as a surface integral:
\begin{eqnarray}
W = R_\odot^2 \, \int \tilde W(t,\theta) \sin \theta \, d\theta \, d\phi
=2 \pi \,  R_\odot^2 \, \int \tau_s(\meanB) \,I(t,\theta) \sin \theta \,d\theta .
\label{M16}
\end{eqnarray}
To determine the function $\tau_s(\meanB)$ we take into account that when the solar activity increases (decreases), the life time of sunspots increases (decreases), so that $\tau_s(\meanB)$ is
\begin{eqnarray}
\tau_s(\meanB)=\tau_\ast \exp \left(C_s \, \partial \meanB/\partial t\right) ,
\label{M17}
\end{eqnarray}
with $C_s= 1.8 \times 10^{-3}$ and $\tau_\ast \, \gamma_{\rm inst} \sim 10$.
Here the non-dimensional rate of the mean magnetic field, $\partial \meanB/\partial t$, is measured in the units $\xi \meanB_{\rm eq} / t_{\rm td}$, and $t_{\rm td}$ is the turbulent magnetic diffusion time. A particular form of the function $\tau_s(\meanB)$ weakly affects the dynamics of the Wolf numbers.

To obtain a rich realistic behaviour of solar activity,
it is necessary to include both algebraic and dynamical quenching.
The reason is that the magnetic part of the $\alpha$ effect
which is proportional to $1/\meanrho$, is located at the upper part
of the convective zone. The dynamical nonlinear equation for
the magnetic part of the $\alpha$ effect can give rich (chaotic) behaviour.
For example, minimum three nonlinear coupling equations
allow appearance of chaos in dynamical system.
Additional algebraic quenching of both, kinetic and magnetic parts
of the $\alpha$ effect allow easily saturate growth of the large-scale
magnetic field \citep{Kleeorin(2003),Kleeorin(2016),zhang2006}.
This quenching describes the feedback reaction of the strong magnetic field
on the fluid motions and should be taken into account in the nonlinear
dynamo model.

\section{Results}

\begin{figure}
\centering
\includegraphics[width=10cm]{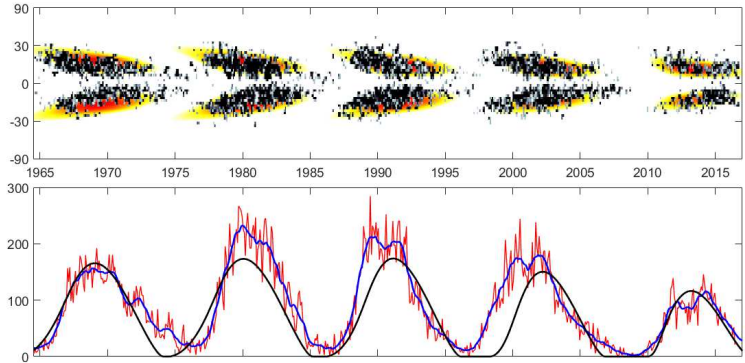}
\caption{\label{Fig1}
Comparison of the results using the dynamo model and observations;
(i) top panel: butterfly diagram of the Wolf number variation rate
$2 \pi \, \sin \theta \,  I(t,\theta)$, the dynamo model (colour) and
the real monthly observational data (black);
(ii) bottom panel: the Wolf numbers, the dynamo model (black),
real observational data (red), and observational data averaged over 13 months (blue).
}
\end{figure}

We solve numerically Eqs.~(\ref{M1}), (\ref{M2}), (\ref{M9}) and~(\ref{M12}).
The parameters of the numerical simulation are as follows: $D=-8450$, $G=1$, $\sigma=3$, $\mu=3$, $\xi=0.1$, $\kappa_{_{T}}=0.1$, $R_\alpha=2$, $T=6.3$, $S_1=0.051$, $S_2=0.95$. We use the following initial conditions: $\meanB_\phi(t=0,\theta)=S_1 \sin\theta + S_2 \sin(2\theta)$ and $\meanAAA(t=0,\theta)=0$.
Comparison of the results using the dynamo model and observations are shown in Fig.~\ref{Fig1}.
In the top panel of Fig.~\ref{Fig1} we show the butterfly diagram of the Wolf number variation rate $2 \pi \, \sin \theta \,  I(t,\theta)$, obtained from the numerical solution of equations of the dynamo model described in Sec.~\ref{model}.
We also compare these results with those obtained from observations.
We use the observed Wolf numbers time series (the real monthly observational data known as the monthly mean total sunspot number, red line).
The data is available in open access from the World Data Center SILSO, Royal Observatory of Belgium, Brussels.
We also show these observational data using a 13 months sliding (or window) averaging of the observed Wolf numbers time series (blue line).
In the bottom panel of Fig.~\ref{Fig1} we show the time evolution of the Wolf numbers based on the dynamo model and observations.
The dynamo model reproduces some of the observed features such as the latitude
distribution of the active regions.

\begin{figure}
\centering
\includegraphics[width=10cm]{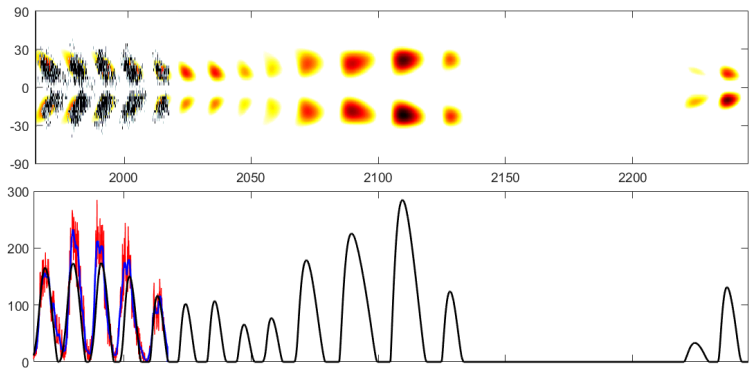}
\caption{\label{Fig2}
The long-term evolution of large-scale magnetic field and the Wolf number time
series; (i) top panel: butterfly diagram of the Wolf number variation rate
$2 \pi \, \sin \theta \,  I(t,\theta)$, the dynamo model (colour) and
the real monthly observational data (black);
(ii) bottom panel: the Wolf numbers, the dynamo model (black),
real observational data (red), and observational data averaged over 13 months (blue).
}
\end{figure}

The long-term evolution of these characteristics is shown in Fig.~\ref{Fig2}.
In particular, the considered dynamo model is able to produce very rich behaviour which includes a decrease of the solar activity during the next half century up to the minimum (similar to Dalton minimum), followed by strong increase of solar activity and the Maunder minimum.
The long-term behavior of the dynamo model
(see Fig.~\ref{Fig2}) shows an example of the behavior of the dynamo model, but not yet
a forecast of solar activity, because no observational data has been 
assimilated into the model here.
Note that the simulation discussed here includes 150 years (14 cycles) of transient period.
The solutions in Figures~\ref{Fig1} and~\ref{Fig2} are shown in later simulated time, 
so any transient is not seen in these figures.

The poleward propagation of the Wolf numbers is observed
in Fig.~\ref{Fig2} (upper panel) starting from $t=2075$ up to $t=2125$.
In the framework of the used simplified dynamo model
there is either equatorward propagation of the sunspot belt
or poleward propagation of the sunspot belt depending on the level
of the magnetic part of the $\alpha$ effect. However,
there are no simultaneously coexisting two branches of the dynamo waves.
To obtain simultaneously coexisting two branches of the dynamo waves,
a two-layer dynamo model with different signs of the differential rotation
can be considered in the dynamo model \citep{belvedere2000}.

Note that sequences of several cycles with equally high correlation to the solar activity between 1965 and present day  can be found with various different sets of system parameters or even different time spans within one particular simulation. To reach the high correlation between simulated and observed data, we compared different characteristics in simulated and observed sunspot cycles (see Figures 9 - 11 in \cite{Kleeorin(2016)}).

\begin{figure}
\centering
\includegraphics[width=14cm]{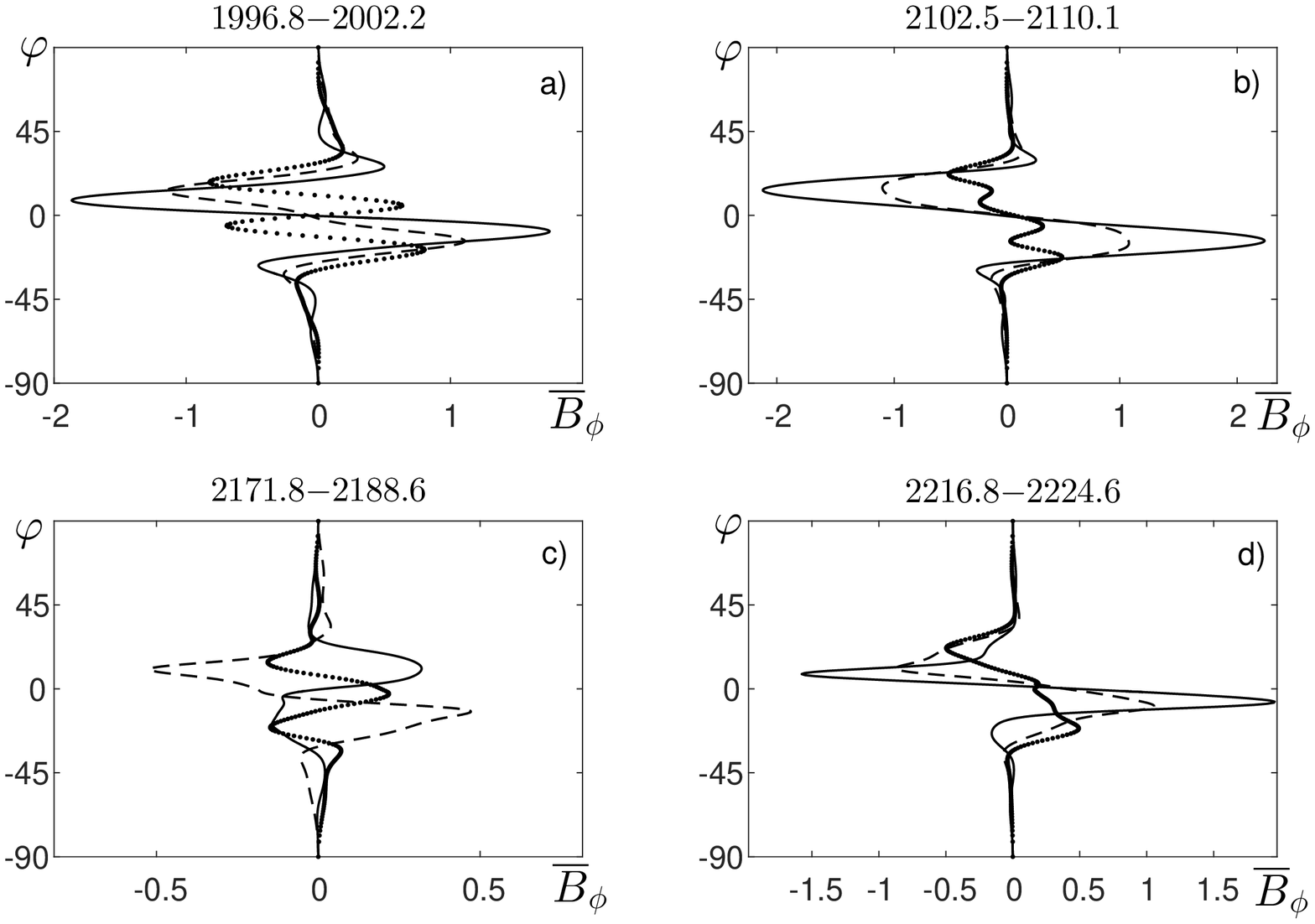}
\caption{\label{Fig3}
The latitude distribution of the toroidal magnetic field $\meanB_\phi(\varphi)$ in different stages of the magnetic field evolution obtained in the dynamo model: no sunspots (dotted line); beginning of the solar cycle (dashed line) and the maximum of solar activity (solid line). The different panels correspond to different epochs: (a) the modern epoch; (b) the future epoch with high solar activity; (c) the possible Maunder minimum like epoch; (d) the epoch after the possible Maunder minimum. The magnetic field is measured in the units of the threshold of the sunspot formation.}
\end{figure}

In Fig.~\ref{Fig3} we show the latitude distribution of the toroidal magnetic field $\meanB_\phi(\varphi)$ in different stages of the magnetic field evolution obtained in the framework of the dynamo model described in Sec.~\ref{model}. This evolution includes lower solar activity without sunspots, the beginning of the new solar cycle and the maximum of solar activity. The different panels correspond to different epochs, e.g., the modern epoch of solar activity; the future epoch with high level of the solar activity; the possible Maunder minimum epoch; and the epoch after the Maunder minimum.
The new cycle in the modern epoch starts in high latitudes and the dynamo waves propagates
to the equator.
On the other hand, in the case of very high level of the solar activity, its maximum reaches very fast at lower latitudes and the dynamo wave propagates to the higher latitudes.
During the possible Maunder minimum epoch, a strong asymmetry between the north and the south hemispheres is observed.

After entering into the Maunder minimum the model again comes back on suddenly
(see Fig.~\ref{Fig2}). Let us discuss what causes the sun to be kicked into and out of the Maunder minimum.
The considered dynamo model, that includes the equations for the poloidal and toroidal mean
magnetic fields, the dynamical equation for the magnetic part of the $\alpha$
effect and the budget equation for the surface density of the Wolf number,
can produce rich behaviour including the Maunder minimum.
The first three nonlinear equations describe
the chaotic behaviour of the large-scale magnetic field.
The last equation mimics the sunspots formation
which takes into account the threshold
for excitation of NEMPI in the large-scale magnetic field.
The latter provides switch on and switch off
of the sunspots formation (see below).

In the upper panel of Fig.~\ref{Fig2} the propagation direction of the dynamo
wave changes after which the cycles disappear for almost a century.
The physics for the disappearance of the sunspots before starting
the Maunder minimum is as follows.
When the mean magnetic field decreases below the
threshold for the large-scale instability (i.e., NEMPI),
the sunspots cannot be formed anymore (see Fig.~\ref{Fig3}c).
Before it happened, the level of magnetic activity
was high (see lower panel of Fig.~\ref{Fig2}),
and polar branch of activity was dominating (see upper panel of Fig.~\ref{Fig2}).
The latter is because the total $\alpha$ effect (that determines the direction of the
dynamo wave propagation) changes sign.
The reason is that when the mean magnetic field becomes strong,
the magnetic part of the $\alpha$ effect can be larger than the
the kinetic part of the $\alpha$ effect, so that the total
$\alpha$ changes sign. We remind that
the magnetic and kinetic parts of the $\alpha$ effect
have opposite signs.

We performed a parameter scan using about $10^3$ runs with different sets of parameters
to find an optimal set of parameters to reach a high level of correlation between
the dynamo model results and observations of the Wolf numbers.
Let us discuss how the variations of the parameters affect the results.
In our previous study \citep{Kleeorin(2016)} we found that
there are two crucial parameters which strongly affect the dynamics
of the nonlinear dynamo system: the dynamo number $D$
and the initial field $B_{\rm init}^{\rm dip}$ for the dipole mode,
determined by the parameter $S_2$.
A proper choice of the initial field $B_{\rm init}^{\rm dip}$ allows
to avoid very long transient regimes to reach the strange attractor.
Comparing the results of the dynamo model with observations, we determine
the correlation between the numerical simulation data
for the Wolf number and the observational data.
To find the maximum correlation between the dynamo model results and
the observed Wolf numbers, the following
parameter scan has been performed: $- 8800 \leq D \leq - 8200$
and $0.85 \leq S_2 \leq 0.95$ (see, e.g., Fig.~12 in \cite{Kleeorin(2016)}).
The maximum correlation is obtained when the parameters are $D=-8450$ and $S_2=0.95$.
The function $D(\sigma)$ determines the region of chaotic behaviour, and
for small $\sigma$ the dynamo system cannot remain inside the region
of the chaotic behaviour.
To find the region of the chaotic behaviour, the following
parameter scan has been performed: $- 10^4 \leq D \leq - 3 \times 10^3$
and $0.3 \leq \sigma \leq 9$ (see, e.g., Fig.~1 in \cite{Kleeorin(2016)}).
The parameter $\mu$ determines the critical dynamo number, $|D_{\rm cr}|$,
for the excitation of the large-scale dynamo instability.
The flux of the magnetic helicity [see Eqs.~(\ref{M8}) and ~(\ref{M9})],
characterised by the parameter $\kappa_{_{T}}$, cannot be very small to
avoid the catastrophic quenching of the $\alpha$ effect. The optimal value
for this parameter is $\kappa_{_{T}} \approx 0.1$.
The variations of the other parameters only weakly affect
the obtained results \citep{Kleeorin(2016)}.

\section{Forecast of solar activity}

Any mean-field dynamo model works on s time-scale that is larger than 1 year.
Indeed, according to the model of solar convective zone by
\cite{spruit1974}, at the bottom
of the convective zone, say at depth $ h_\ast \sim 2 \times 10^{10}$ cm,
the magnetic Reynolds number ${\rm Rm} \sim 2 \times 10^9$,
the turbulent velocity $u_0 \sim 2 \times 10^3 $ cm s$^{-1}$,
the turbulent scale $\ell_0 \sim 8 \times 10^9$ cm,
so the turn over time of turbulent eddies $\ell_0/u_0 \sim 4 \times 10^6$ s
(that is 0.13 of years). This implies that the mean-field time
(the characteristic time of the mean fields variations)
should be at least one order of magnitude larger than the turn over time,
i.e., about 1 year.
This refers	to a sufficient separation of
temporal scales in which case memory effects can be neglected.
This implies that a mean-field dynamo model cannot provide the forecast of the solar activity on a time-scale of several months.
To predict the solar activity on a short time-scale, additional methods should be used for the forecast of the solar activity.

To predict solar activity on the time scale of one month we use a method based on a combination of the numerical solution of the nonlinear mean-field dynamo equations and the artificial neural network approach \citep{Hagan(2016),Conway(1998),Fessant(1996)}.
A simplified version of the artificial neural networks method to forecast the solar activity has been used before \citep{Conway(1998),Fessant(1996)}. However, this method has not been combined with the advanced mean-field approach based on the nonlinear dynamo models, and the used network scheme has not been stable resulting in a systematical increase of errors.
The recent developments in the field of artificial neural networks and the increased computational capabilities of the computers allow to combine the simulations of the nonlinear mean-field dynamo model with the artificial neural network forecast scheme.

To apply this approach, we use the initial simulations of the Wolf numbers $W_i^{\rm model}$, based on the dynamo model described in Sec.~\ref{model}, as the basis for the forecast and as the exogenous input in the neural network scheme. Another input is the data $W_i^{\rm obs}$
obtained from observations. To perform the forecast $W_i^{\rm forecast}$ for the next half a year or longer, we adopt an autoregressive scheme with unknown coefficients (determined during the learning procedure):
\begin{eqnarray}
W_i^{\rm forecast} = {f_{\rm out}}\left( {{\bm{K} \, \bm{w}} + {\bm{c}}} \right) ,
\label{T1}
\end{eqnarray}
where $f_{\rm out}(x)$ is a linear function of an outer layer of neurons, ${\bm{c}}$ is the vector of biases; ${\bm{K}}$ is the weight matrix of neurons; ${\bm{w}}$ is the inputs vector, containing observations $W_i^{\rm obs}$ and model estimations $W_i^{\rm model}$. To estimate the weight matrix ${\bm{K}}$ one needs to minimize the error between the deviations of the forecast data $W_i^{\rm forecast}$ from the observational data $W_i^{\rm obs}$.

Equation~(\ref{T1}) describes a simple ''one-layer artificial neural network''. However, for our task it is required to use a more complex scheme, e.g., a ''two-layer artificial neural network'' type of a recurrent dynamic nonlinear autoregressive network, with feedback connections enclosing two layers of the network, defined by the following equation:
\begin{eqnarray}
W_i^{\rm forecast} = {f_{\rm out}}\left[ {{{\bm{K}}_{2}}\, {f_{\rm hidden}}\left( {{{\bm{K}}_{1}}\, {\bm{w}} + {{\bm{c}}_1}} \right) + {{\bm{c}}_2}} \right],
\label{T2}
\end{eqnarray}
where $f_{\rm hidden}(x) = [1 + \exp(-x)]^{-1}$ is a function of a hidden layer of neurons,
${\bm{K}}_{1}$ is the weight matrix $24 \times 8$ of a hidden layer neurons,
${\bm{K}}_{2}$ is the weight matrix $1 \times 24$ of an outer layer neurons,
${{\bm{c}}_1}$ and ${{\bm{c}}_2}$ are the corresponding bias vectors,
$\bm{w}$ is the input vector $8 \times 1$ consisting of 4 prior observations
$\begin{array}{*{20}{c}} {W_{i - 1}^{{\rm{obs}}}}, &  \cdots \,, & {W_{i - 4}^{{\rm{obs}}}}  \\ \end{array}$ and 4 corresponding model estimations
$\begin{array}{*{20}{c}} {W_{i - 1}^{{\rm{model}}}}, &  \cdots \,, &  {W_{i - 4}^{{\rm{model}}}}  \\ \end{array}$.

\begin{figure}
\centering
\includegraphics[width=13cm]{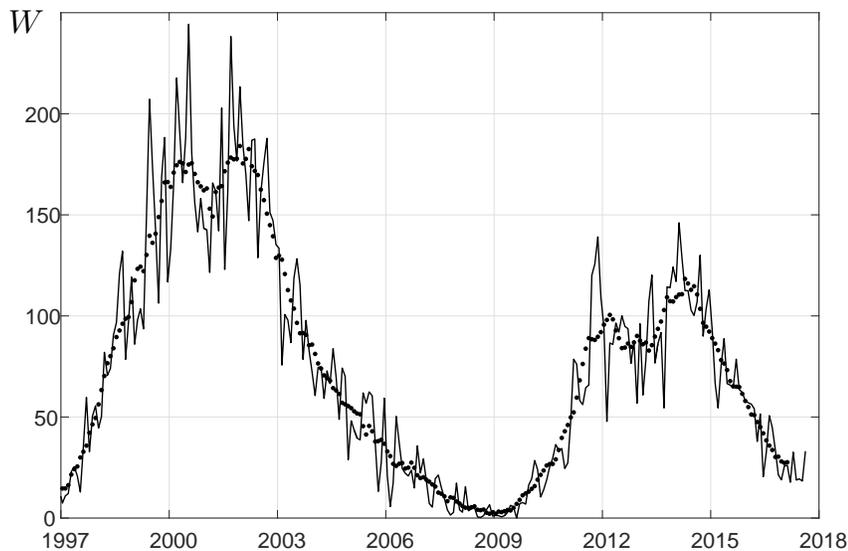}
\caption{\label{Fig4}
Comparison of the results of the one-month forecast of the solar activity (filled circles) with the observed Wolf numbers (solid line).}
\end{figure}

Equation~(\ref{T2}) provides more stable, complex, adaptable and adjustable forecast than Eq.~(\ref{T1}), e.g., in the presence of noise. The learning procedure by Bayesian regularization back-propagation has been based on epignose using the data of the Wolf numbers from 19-20 cycles, while 21 cycle has been used for the validation process.
The input data of the Wolf numbers for the neural network consists of two parts: the prior real observations (e.g., red line in Fig.~\ref{Fig1}) and the dynamo model estimations at the same instant (e.g., solid line in Fig.~\ref{Fig1}). The output of this neural network is the forecasted monthly Wolf number.
Note that we do not use the artificial neural network for any type of optimisation or parameter estimation for the initial model. We have already done this in our previous study \citep{Kleeorin(2016)}.
During the learning procedure in the artificial neural network, we minimize the error between forecast and actual observations not in every instant separately, but over the whole cycle.
We stress that the model output yields an initial forecast, i.e., a basis for the final forecast. The artificial neural network serves here as a forecast correction scheme for the simulated sunspots. The correction is done by means of observational data and the model outputs.

\begin{figure}
\centering
\includegraphics[width=13cm]{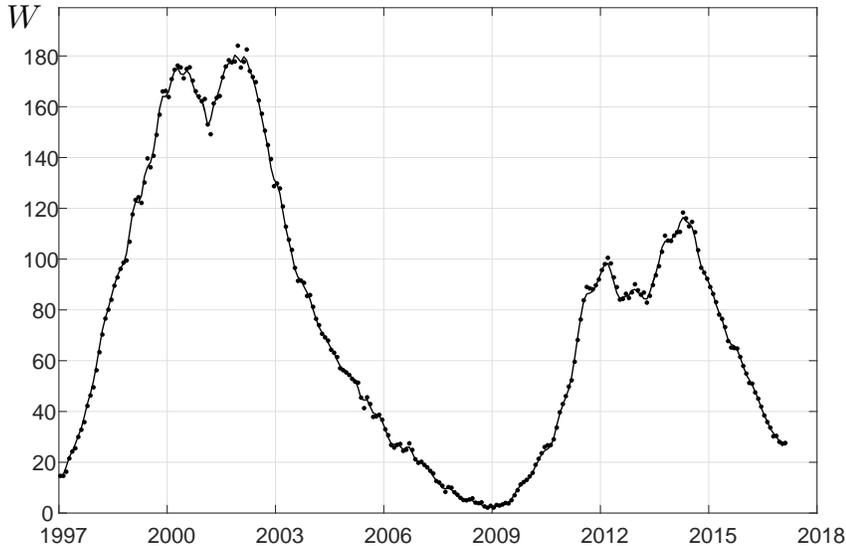}
\caption{\label{Fig5}
Comparison of the results of the one-month forecast of the solar activity (filled circles) with the observed Wolf numbers averaged over 13 month (solid line).}
\end{figure}

\begin{figure}
\centering
\includegraphics[width=13cm]{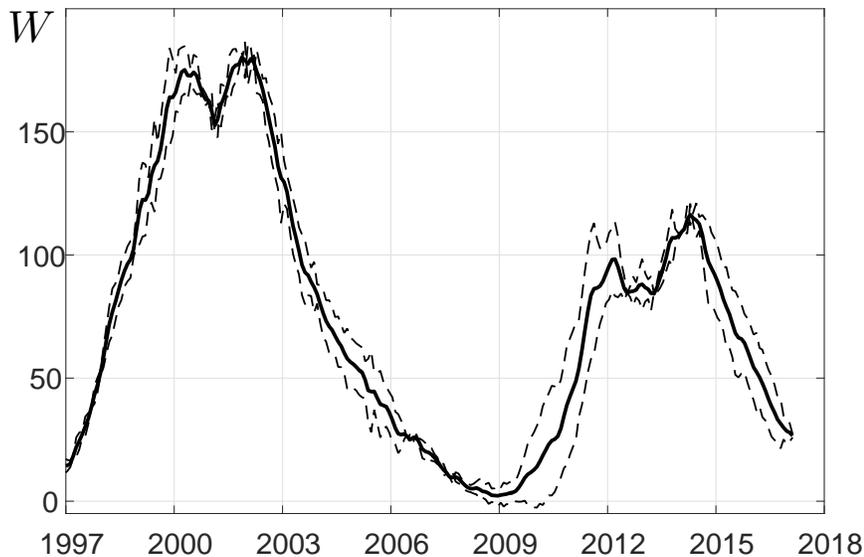}
\caption{\label{Fig6}
The confidence interval (dashed lines) of the one-month forecast of the solar activity compared with the observed Wolf numbers averaged over 13 month (solid line).}
\end{figure}

The obtained results of the forecast of the solar activity are presented below.
In Fig.~\ref{Fig4} we show a comparison of the results of the one-month
forecast of the solar activity based on the described method with the observed Wolf numbers,
while in Fig.~\ref{Fig5} we show the same comparison but with the data of the observed Wolf numbers averaged over 13 month.
The sliding (or window) averaging of the observed Wolf numbers time series
has been used here.
In Fig.~\ref{Fig4} we use the observed Wolf numbers time series with 1 month averaging time,
while in Fig. 5 we use the observed Wolf numbers time series with 13 month averaging time.
Sampling time is exactly one month in both cases.
For the latter case
we show in Fig.~\ref{Fig6} also the confidence interval for this forecast.
Note that for the cycle with a higher activity the thickness of the confidence interval is less. The decrease of the confidence interval is also observed in the phase of the increasing of the solar activity (see Fig.~\ref{Fig6}).
These figures demonstrate a good agreement between the forecast and observation of the solar activity.

We would like to stress that only the combination of the numerical solution of the nonlinear mean-field dynamo equations and the neural network yields good agreement between the forecast and observation. Using only the neural network without the mean-field solution provides reasonable agreement just for 5 years because there is no a long-term memory in the magnetic field evolution.

\section{Conclusions}

To predict the solar activity, we apply the nonlinear mean-field dynamo model with algebraic and dynamic nonlinearities of the alpha effect and with a budget equation for the dynamics of Wolf numbers.
We use a simplified axi-symmetric dynamo model that allows us to obtain very long time series of Wolf numbers.
This dynamo model demonstrates very rich behaviour, reproducing the observed evolution
of the magnetic activity during past three centuries.
We forecast the solar activity on the time scale of one month adopting a method consisting in a combination of the numerical solution of the nonlinear mean-field dynamo equations and
the artificial neural network.
A comparison between the forecast of solar activity and the observed Wolf numbers shows a good agreement, which is achieved due to the combined effect of the nonlinear mean-field dynamo model and the artificial neural network technique. Without the mean-field model
it is impossible to get a reasonable agreement in a long-term evolution.

\bigskip

\begin{acknowledgements}
This work was supported in part by the Research Council of Norway
under the FRINATEK (grant No. 231444).
The authors acknowledge the hospitality of NORDITA, Ural Federal University
and the Kavli Institute for Theoretical Physics in Santa Barbara.
\end{acknowledgements}

\newpage

\bibliographystyle{jpp}

\bibliography{dynamo-solar-activity-JPP}

\end{document}